\documentclass[journal,comsoc]{IEEEtran}

\usepackage{xspace,amsmath,amssymb,amsfonts,epsfig,syntonly}
\usepackage{cite,bm,color,url,textcomp,empheq,boxedminipage}
\usepackage{algorithmicx,algorithm,algpseudocode}
\usepackage{epstopdf}
\usepackage{empheq}
\usepackage{graphicx,graphics,subfigure}  % Written by David Carlisle and Sebastian Rahtz
\usepackage{multirow,multicol}
\usepackage{psfrag}    % Written by Craig Barratt, Michael C. Grant,
\usepackage{stfloats}

\newtheorem{lemma}{\underline{Lemma}}[section]

\newtheorem{proposition}{\underline{Proposition}}[section]

\newtheorem{remark}{\underline{Remark}}[section]

\newcommand{\mv}[1]{\mbox{\boldmath{$ #1 $}}}

\long\def\symbolfootnote[#1]#2{\begingroup
\def\thefootnote{\fnsymbol{footnote}}
\footnote[#1]{#2}\endgroup}

\psfull

\usepackage[T1]{fontenc}% optional T1 font encoding
\usepackage{ifpdf}
 \ifpdf
 \else
 \fi

\begin{document}
\title{Joint Offloading and Computing Optimization in Wireless Powered Mobile-Edge Computing~Systems}

\author{Feng~Wang,~\IEEEmembership{Member,~IEEE,}
        Jie~Xu,~\IEEEmembership{Member,~IEEE,}
        Xin~Wang,~\IEEEmembership{Senior Member,~IEEE,}
        and~Shuguang~Cui,~\IEEEmembership{Fellow,~IEEE}
\thanks{Manuscript received May 26, 2017; revised October 16, 2017; accepted December 12, 2017. This work was in part supported by the National Key Research and Development Program of China grant no. 2017YFB0403402 and the National Natural Science Foundation of China grant no. 61671154, by DoD with grant HDTRA1-13-1-0029, by grant NSFC-61328102/61629101, by Shenzhen Fundamental Research Fund under Grant No. KQTD2015033114415450, and by NSF with grants DMS-1622433, AST-1547436, ECCS-1508051/1659025, and CNS-1343155. Part of this paper was presented in the IEEE International Conference on Communications, Paris, France, May 21--25, 2017 \cite{Conf2017}. The associate editor coordinating the review of this paper and approving it for publication was T. Melodia. \emph{(Corresponding author: Jie Xu.)} }

\thanks{F. Wang and J. Xu are with the School of Information Engineering, Guangdong University of Technology, Guangzhou, China (e-mail: fengwang13@gdut.edu.cn, jiexu@gdut.edu.cn).}

\thanks{X. Wang is with the Key Laboratory for Information Science of Electromagnetic Waves (MoE), the Shanghai Institute for Advanced Communication and Data Science, the Department of Communication Science and Engineering, Fudan University, Shanghai, China (e-mail: xwang11@fudan.edu.cn).}

\thanks{S. Cui is with the Department of Electrical and Computer Engineering, University of California, Davis, CA, 95616, USA, and also affiliated with Shenzhen Research Institute of Big Data (e-mail: sgcui@ucdavis.edu).}}
%S. Cui is with the Department of Electrical and Computer Engineering, University of California, Davis, CA, 95616. He is also affiliated with Shenzhen Research Institute of Big Data.

%The work of S. Cui was supported in part by DoD with grant HDTRA1-13-1-0029, by grant NSFC-61328102/61629101, by Shenzhen Fundamental Research Fund under Grant No. KQTD2015033114415450, and by NSF with grants DMS-1622433, AST-1547436, ECCS-1508051/1659025, and CNS-1343155.

%{Shell \MakeLowercase{\textit{et al.}}: Bare Demo of IEEEtran.cls for IEEE Communications Society Journals}


\maketitle

\begin{abstract}
Mobile-edge computing (MEC) and wireless power transfer (WPT) have been recognized as promising techniques in the Internet of Things (IoT) era to provide massive low-power wireless devices with enhanced computation capability and sustainable energy supply. In this paper, we propose a unified MEC-WPT design by considering a wireless powered multiuser MEC system, where a multi-antenna access point (AP) (integrated with an MEC server) broadcasts wireless power to charge multiple users and each user node relies on the harvested energy to execute computation tasks. With MEC, these users can execute their respective tasks locally by themselves or offload all or part of them to the AP based on a time division multiple access (TDMA) protocol. Building on the proposed model, we develop an innovative framework to improve the MEC performance, by jointly optimizing the energy transmit beamforming at the AP, the central processing unit (CPU) frequencies and the numbers of offloaded bits at the users, as well as the time allocation among users. Under this framework, we address a practical scenario where latency-limited computation is required. In this case, we develop an optimal resource allocation scheme that minimizes the AP's total energy consumption subject to the users' individual computation latency constraints. Leveraging the state-of-the-art optimization techniques, we derive the optimal solution in a semi-closed form. Numerical results demonstrate the merits of the proposed design over alternative benchmark schemes.
\end{abstract}


\IEEEpeerreviewmaketitle

\section{Introduction}

 The recent advancement of Internet of Things (IoT) has motivated various new applications (e.g., autonomous driving, virtual reality, augmented reality, and tele-surgery) to provide real-time machine-to-machine and machine-to-human interactions~\cite{Chiang16}. These emerging latency-sensitive applications critically rely on the real-time communication and computation of massive wireless devices. For example, smart wireless sensors in IoT networks may need to perceive the physical environment and then use the built-in computation resources to preprocess the sensed data in real time before sending it to the access point (AP)~\cite{Chiang16}. As extensive existing works focus on improving their communication performance, how to provide these devices with enhanced computation capability is a crucial yet challenging task to be tackled, especially when they are of small size and low power. To resolve this issue, mobile-edge computing (MEC) has emerged as a promising technique by providing cloud-like computing at the edge of mobile networks via integrated MEC servers at wireless APs and base stations (BSs)\cite{Bar14,Mao17}. Leveraging MEC, resource-limited wireless devices can offload their computation tasks to APs/BSs; then the integrated MEC servers can compute these tasks remotely. In general, the computation offloading can be implemented in two ways, namely {\em binary} and {\em partial} offloading\cite{Mao17}. In the binary offloading case, the computation task is not partitionable and should be offloaded as a whole. In the partial offloading case, the task can be partitioned into two parts, and only one of them is offloaded. The MEC technique facilitates the real-time implementation of computation-extensive tasks at massive low-power devices, and thus has attracted growing research interests in both academia and industry \cite{ETSI,ThinkAir12,MAUI10,Bar14,Mao17}.

On the other hand, how to provide sustainable and cost-effective energy supply to massive computation-heavy devices is another challenge facing IoT. Radio-frequency (RF) signal based wireless power transfer (WPT) provides a viable solution by deploying dedicated energy transmitters to broadcast energy wirelessly\cite{Xu14_1bit}. Recently, emerging wireless powered communication networks (WPCNs) and simultaneous wireless information and power transfer (SWIPT) paradigms have been proposed to achieve ubiquitous wireless communications in a self-sustainable way \cite{Bi15,Li15,Xu14,Feng15}, where WPT and wireless communications are combined into a joint design. In order to improve the WPT efficiency from the energy transmitter to one or more energy receivers, {\em transmit energy beamforming} has been proposed as a promising solution by deploying multiple antennas at energy transmitters \cite{Zhang13}. By properly adjusting the transmit beamforming vectors, energy transmitters can concentrate the radiative energy towards the intended receivers for efficient WPT. Motivated by these approaches, it is expected that the transmit energy beamforming-enabled WPT can also play an important role in facilitating self-sustainable computing for a large number of IoT devices.

To explore benefits of both MEC and WPT in ubiquitous computing, this paper develops a joint MEC-WPT design by considering a wireless powered {\em multiuser} MEC system that consists of a multi-antenna AP and multiple single-antenna users. The AP employs energy transmit beamforming to simultaneously charge the users, and each user relies on its harvested energy to execute the respective computation task. Suppose that partial offloading is allowed such that each user can arbitrarily partition the computation task into two \emph{independent} parts for local computing and offloading, respectively. Furthermore, we assume that the downlink WPT and the uplink wireless communication (for computation offloading) are operated simultaneously over orthogonal frequency bands.\footnote{The wireless energy harvesting in the downlink and the information transmission (or offloading) in the uplink can be performed simultaneously over orthogonal frequency bands in one single antenna with a duplexer, as commonly used in conventional frequency-division-duplexing (FDD) wireless communication transceivers.} In addition, a time division multiple access (TDMA) protocol is employed to coordinate computation offloading, where different users offload their respective tasks to the AP over orthogonal time slots. The main results of this paper are summarized as follows.

\begin{itemize}
\item To improve the performance of such a wireless powered multiuser MEC system, we develop an innovative design framework by jointly optimizing the energy transmit beamforming at the AP, the central processing unit (CPU) frequencies\footnote{The term {\em CPU} generally refers to the processing unit and control unit at each user that takes charge of the local computing of computation tasks. The CPU frequency, i.e., the frequency of the CPU's clock pulses, determines the rate at which a CPU executes instructions.} and the numbers of offloaded bits at the users, as well as the offloading time allocation among users. Note that the number of offloaded bits at each user corresponds to the multiplication of the offloading rate and allocated offloading time in this block.

\item Targeting an energy-efficient wireless-powered MEC design, we minimize the AP's total energy consumption subject to the users' individual computation latency constraints. Leveraging the state-of-the-art optimization techniques, we obtain the optimal solution in a semi-closed form. It is revealed that at the optimal solution, the number of locally computed bits at each user should be strictly positive; i.e., it is always beneficial for each user to leave certain bits for local computing. It is also shown that the optimal offloading rate (and equivalent transmit power) at each user critically depends on the channel power gain and the circuit power.

\item Extensive numerical results are provided to gauge the performance of the proposed designs with joint WPT, local computing, and offloading (i.e., task partition per user and offloading time allocation among the users) optimization, over benchmark schemes without such a joint consideration. It is shown that the proposed design can significantly reduce the energy consumption of the wireless powered MEC systems.
\end{itemize}

\subsection{Related Works}
Transmit energy beamforming enabled WPT has been extensively studied in the literature (see, e.g., \cite{Zhang13,Xu14_1bit,Xu14,Xu16_general,Feng15,Li15,Ng14,Bi15,Tim14,Feng17,Zeng15,Valenta14,Zeng17,Bosh15,Clerckx16,Lee2017} and references therein). By considering a linear energy harvesting (EH) model, various prior works have investigated the optimal design of energy beamforming under different setups with SWIPT, e.g., in two-user multi-input multi-output (MIMO) systems \cite{Zhang13}, secrecy communications systems \cite{Ng14}, multi-input single-output (MISO) interference channels \cite{Tim14}, and multiuser MISO downlink channels \cite{Xu14,Feng17,Feng15}. Furthermore, some recent works investigated the transmit power allocation \cite{Bosh15} and the transmit waveform optimization \cite{Clerckx16} for WPT by taking into account the nonlinear nature of the rectifier in EH \cite{Valenta14,Zeng17}. In addition, the benefit of energy beamforming crucially relies on the channel state information (CSI) known at the transmitter. The reverse-link channel training \cite{Zeng15} and the energy measurement and feedback methods \cite{Xu14_1bit,Xu16_general} were proposed in WPT systems for the energy transmitter to practically learn the CSI to users. Furthermore, \cite{Lee2017} developed a distributed energy beamforming system for multiple energy transmitters to charge multiple energy receivers simultaneously, with the help of the energy measurement and feedback.

On the other hand, several existing works\cite{Huang16,Liu16,Mun15,Huang12,Chen16,Wang16,X_Chen16,Liu13,Sar15} investigated the energy-efficient multiuser MEC design, where each user is powered by fixed energy sources such as battery, and the objective is to minimize the energy consumption at the users via joint computing and offloading optimization at the demand side. For example, \cite{Liu13} provided an overview on the applications and challenges of computation offloading. \cite{Liu16} and \cite{Huang12} investigated the dynamic offloading for MEC systems based on the techniques of Markov decision process and Lyapunov optimization, respectively. \cite{Mun15,Wang16} considered the joint computation and communication resource allocation in single-user MEC systems, and such designs were extended to multiuser MEC systems in \cite{Chen16,X_Chen16,Huang16,Sar15}. Different from these prior works that studied WPT and MEC separately, this paper pursues a joint MEC-WPT design in a wireless powered multiuser MEC system, by jointly optimizing the WPT supply at the AP, as well as the local computing and offloading demands at the users.

It is worth noting that a prior work \cite{You16} considered the wireless powered {\em single-user} MEC system with binary offloading, where the user aims to maximize the probability of successful computation, by deciding whether a task should be fully offloaded or not, subject to the computation latency constraint. By contrast, this paper considers a more general case with more than one user, and allows for more flexible partial offloading to improve the system performance in terms of the energy efficiency (i.e., minimizing the total energy consumption at the AP including the radiated energy for WPT and the energy for computing the offloaded tasks).

 The remainder of the paper is organized as follows. Section II presents the system model. Section III formulates the computation latency constrained energy consumption minimization problem, and develops an efficient algorithm to obtain a well-structured optimal solution. Section IV provides numerical results to demonstrate the merits of the proposed design. Finally, Section V concludes this paper.

{\em Notations:} Boldface letters refer to vectors (lower case) or matrices (upper case). For a square matrix $\boldsymbol{S}$, ${\rm tr}(\boldsymbol{S})$ denotes its trace, while $\boldsymbol{S}\succeq \boldsymbol{0}$ means that $\boldsymbol{S}$ is positive semidefinite. For an arbitrary-size matrix $\boldsymbol{M}$, ${\rm rank}(\boldsymbol{M})$, $\boldsymbol{M}^\dagger$, and $\boldsymbol{M}^H$ denote its rank, transpose, and conjugate transpose, respectively. $\boldsymbol{I}$ and $\boldsymbol{0}$ denote an identity matrix and an all-zero vector/matrix, respectively, with appropriate dimensions. $\mathbb{C}^{x\times y}$ denotes the space of $x\times y$ complex matrices; $\mathbb{R}$ denotes the set of real numbers. $\mathbb{E}[\cdot]$ denotes the statistical expectation. $\|\boldsymbol{x}\|$ denotes the Euclidean norm of a vector $\boldsymbol{x}$, $|z|$ denotes the magnitude of a complex number $z$, and $[x]^+ \triangleq \max(x,0)$.

\section{System Model}

\begin{figure}[!t]
  \centering
  \includegraphics[width = 3.5in]{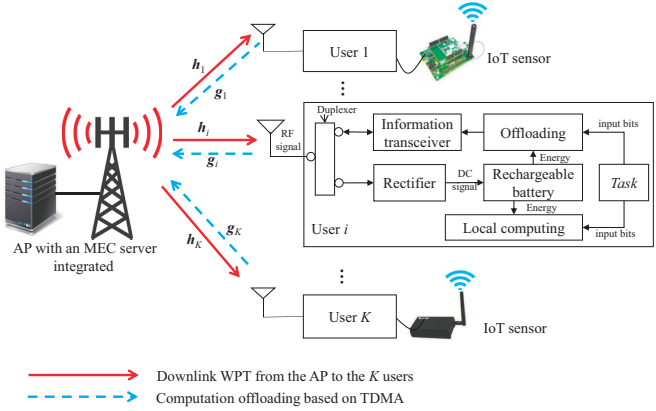}\\
  \caption{A wireless powered multiuser MEC system with WPT in the downlink and computation offloading in the uplink.} \label{fig.SysMod1}
\end{figure}

As shown in Fig.~\ref{fig.SysMod1}, we consider a wireless powered multiuser MEC system consisting of an $N$-antenna AP (integrated with an MEC server) and a set ${\cal K}\triangleq\{1,\ldots,K\}$ of single-antenna users. In this system, the AP employs RF signal based energy transmit beamforming to charge the $K$ users. Each user $i\in{\cal K}$ utilizes the harvested energy to execute its computation task through local computing and offloading. Suppose that the downlink WPT from the AP to the users and the uplink computation offloading are operated simultaneously over orthogonal frequency bands, and the uplink for computation offloading and the downlink for computation result downloading are operated over the same frequency band. Assume a block-based model, and we focus on one particular block with length $T$. Here, $T$ is chosen to be no larger than the latency of the MEC application and also no larger than the channel coherence time, such that the wireless channels remain unchanged during this block. For simplifying the analysis and better capturing the AP's transmission energy for computation offloading, we assume that the AP perfectly knows the CSI from/to the $K$ users,\footnote{When the CSI at the AP is not perfect (e.g., subject to some CSI estimation errors), the WPT and MEC performance may degrade. In this case, robust optimization techniques (see, e.g., \cite{Ng14,Feng15}) may be applied to obtain the energy beamforming vectors. However, the imperfect CSI scenario is out of scope of this paper.} as well as their computation requirements. In accordance with such information, the AP coordinates the downlink WPT, the computation offloading, and the local computing for the $K$ users.

\subsection{Energy Transmit Beamforming from AP to Users}
 Let $\boldsymbol{s}\in\mathbb{C}^{N\times 1}$ denote the energy-bearing transmit signal by the AP, which is assumed to be a random signal with its power spectral density satisfying certain regulations on RF radiation \cite{Xu14_1bit}. Let $\boldsymbol{Q}\triangleq \mathbb{E}[\boldsymbol{s}\boldsymbol{s}^H]\succeq {\bm 0}$ denote the energy transmit covariance matrix and $\mathbb{E}[\|{\bm s}\|^2]={\rm tr}({\bm Q})$ the transmit power at the AP. In general, the AP can use multiple energy beams to deliver the wireless energy, i.e., $\boldsymbol{Q}$ can be of any rank. Supposing $r={\rm rank}(\boldsymbol{Q})\leq N$, then a total of $r$ energy beams can be obtained via the eigenvalue decomposition (EVD) of $\boldsymbol{Q}$~\cite{Xu14_1bit}. Let $\boldsymbol{h}_i\in\mathbb{C}^{N\times 1}$ denote the channel vector from the AP to user $i\in {\cal K}$, and define $\boldsymbol{H}_i\triangleq \boldsymbol{h}_i\boldsymbol{h}_i^H$, $\forall i\in{\cal K}$. Accordingly, the received RF power at each user $i\in{\cal K}$ is given by $|\boldsymbol{h}_i^H\boldsymbol{s}|^2$. In order to harvest such energy, each user $i$ first converts the received RF signal into a direct current (DC) signal by a rectifier and then stores the energy of the DC signal in its chargeable battery (cf. Fig.~1). Note that the harvested DC power is generally nonlinear with respect to the received RF power \cite{Valenta14}, due to the nonlinear devices such as the diodes and diode-connected transistors. Moreover, the nonlinear RF-to-DC conversion greatly depends on the input power level and the transmit waveform. In the literature, there have been a handful of recent works on analytic nonlinear EH models, which characterize such nonlinear relations between the harvested DC power and the input RF power \cite{Bosh15} or transmit waveform \cite{Clerckx16}. However, there still lacks a generic EH model that captures all practical issues\cite{Zeng17}. Therefore, for simplicity, we assume that the input RF power is within the linear regime of the rectifier, and consider a linear EH model which has been commonly adopted in the WPT literature [9]--[12], [14]--[22]. Accordingly, the harvested energy amount by user~$i$ over this time block is
\begin{align}\label{eq.energy_harvested}
E_i =  T\zeta \mathbb{E}\left[\left| \boldsymbol{h}^H_i\boldsymbol{s} \right|^2\right] = T\zeta{\rm tr}(\boldsymbol{Q}\boldsymbol{H}_i),
\end{align}
where $0<\zeta\leq 1$ denotes the constant EH efficiency per user. The harvested energy $E_i$ is used by user $i$ for both computation offloading and local computing.

\subsection{Energy Consumption at Users for Computation}
For each user $i\in{\cal K}$, the computation task with $R_i>0$ computation input bits is partitioned into two parts with $\ell_i\geq 0$ and $q_i\geq 0$ bits, which are offloaded to the MEC server at the AP or locally computed, respectively.\footnote{Each input bit can be treated as the smallest task unit, which includes the needed program codes and input parameters.} We assume that such a partition does not incur additional computation input bits, i.e., $R_i=\ell_i+q_i$, $\forall i\in{\cal K}$.

\subsubsection{Computation Offloading from Users to the AP}
\begin{figure}[!t]
  \centering
  \includegraphics[width = 3.5in]{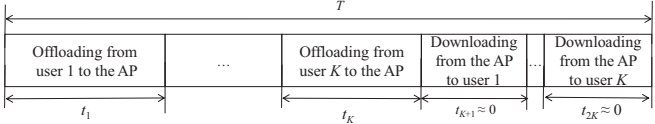}\\
  \caption{The TDMA protocol for multiuser computation offloading.} \label{fig.TDMA}
\end{figure}
In order for the $K$ users to offload their respective bits to the AP for computation, we adopt a TDMA protocol without interference as shown in Fig.~\ref{fig.TDMA}, where the block is divided into $2K$ time slots each with a length of $t_i$, $\forall i\in\{1,\ldots,2K\}$. In the first $K$ time slots, the $K$ users offload their computation bits to the AP one by one. After the MEC server executes the computation tasks on behalf of these users, the AP sends the computation results to the $K$ users in the last $K$ time slots. Due to the sufficient CPU capability at the MEC server, the computation time consumed at the MEC server are relatively small and negligible. Therefore, we assume that the users can download the computation results immediately after the first $K$ offloading time slots. Furthermore, as the AP is usually with high transmit power and the computed results are usually of small size, we ignore the downloading time, i.e., $t_i\approx 0$, $\forall i\in\{K+1,\ldots,2K\}$, and also ignore the energy consumption for transmitting and receiving the computation results in this paper.

For computation offloading in time slot $i$, let $\boldsymbol{g}_i\in \mathbb{C}^{N\times 1}$ denote the uplink channel vector from user $i$ to the AP and $p_i$ the transmit power for offloading. Assume further that the AP employs the maximum ratio combining (MRC) receiver to decode the information. The achievable offloading rate (in bits/sec) for user $i$ is given by
\begin{align}\label{eq.rate}
r_i = B \log_2\left(1+\frac{p_i\tilde{g}_i}{\Gamma\sigma^2}\right),~~\forall i\in{\cal K},
\end{align}
where $B$ denotes the spectrum bandwidth, $\tilde{g}_i \triangleq \|\boldsymbol{g}_i\|^2$ denotes the effective channel power gain from user~$i$ to the AP, $\sigma^2$ is the noise power at the receiver of the AP, and $\Gamma\geq 1$ is a constant accounting for the gap from the channel capacity due to a practical coding and modulation scheme. For simplicity, $\Gamma=1$ is assumed throughout this paper. As a result, the number of offloaded bits $\ell_i$ by user $i$ to the AP can be expressed as
\begin{align}\label{eq.ell}
\ell_i = r_i t_i,~~\forall i\in {\cal K}.
\end{align}

Computation offloading incurs energy consumption at both the $K$ users and the AP. Per user $i\in{\cal K}$, in addition to the transmit power $p_i$, a constant circuit power $p_{{\rm c},i}$ (by the digital-to-analog converter (DAC), filter, etc.) is consumed. The offloading energy consumption at user $i$ is then $E_{{\rm offl},i}=(p_i+p_{{\rm c},i})t_i$. With simple manipulations based on (\ref{eq.rate}) and (\ref{eq.ell}), the transmit power $p_i$ can be expressed as $p_i = \frac{1}{\tilde{g}_i}\beta\left(\frac{\ell_i}{t_i}\right)$,
where $\beta(x) \triangleq \sigma^2(2^{\frac{x}{B}}-1)$ is a monotonically increasing and convex function with respect to $x$.\footnote{Note that to avoid the issue of dividing by zero, we define $\beta\left(\frac{\ell_i}{t_i}\right)=0$ when either $\ell_i=0$ or $t_i=0$ holds.} Hence, the offloading energy consumption at user $i$ is
\begin{align}\label{eq.energy_offl}
E_{\text{offl},i} = \frac{t_i}{\tilde{g}_i}\beta\left(\frac{\ell_i}{t_i}\right) +p_{c,i} t_i.
\end{align}

\begin{remark}
Note that in practice, in order for the AP to acquire the CSI (to the $K$ users) for the energy beamforming design, each user needs to consume a certain amount of energy (e.g., for CSI feedback), and there generally exists a tradeoff between such energy consumption at the users versus the CSI accuracy at the AP. However, with the technical advancements, the user's feedback overhead for CSI acquisition could be made very small. Specifically, there are generally three types of CSI acquisition methods in the literature, namely the channel estimation and feedback \cite{Love08}, reverse-link training based on the channel reciprocity \cite{Zeng15}, and energy measurement and feedback \cite{Xu14_1bit,Xu16_general}. In the energy measurement and feedback method \cite{Xu14_1bit}, each user only needs to measure its harvested energy level over each block and send one feedback bit to the AP per block; based on the feedback bits, the AP can sequentially improve the accuracy of CSI estimation; such a one-bit feedback is negligible when compared to the user reverse-link traffic for task offloading. Thus it is practically reasonable to ignore the feedback overhead and energy consumption at each user.
\end{remark}

As for the AP, the energy is mainly consumed for executing the offloaded computation tasks and transmitting the computation results back to the users\cite{Bar14}. As the AP and its integrated MEC server generally have sufficient communication and computation capacities\footnote{In the case when the MEC server's computing capacity is limited, the computation offloading protocol needs to be redesigned, by taking into account the computation time at the MEC server as well as the computation resource sharing among these different users. Under such a scenario, how to jointly design the WPT and MEC optimally is out of the scope of this paper. It is an interesting direction to pursue in the future work.}, it can adopt a large transmit power (accordingly high communication rate) and a high constant CPU frequency to minimize the latency. In this case, the AP's energy consumption is generally proportional to the totally offloaded bits $\sum_{i=1}^K \ell_i$ from the $K$ users. Therefore, we adopt a simplified linear energy consumption model for the computation at the AP as
\begin{equation}\label{eq.linear_energy_AP}
E_{\text{MEC}} = \alpha \sum_{i=1}^K \ell_i,
\end{equation}
where $\alpha$ denotes the energy consumption per offloaded bit at the AP. In practice, $\alpha$ depends on the transceiver structure of the AP, the chip structure of the MEC server, and its operated CPU frequencies, etc.\cite{Bar14}.

\subsubsection{Local Computing at Users}
 Consider next the local computing for executing $q_i$ input bits at each user $i\in{\cal K}$. Let $C_i$ denote the number of CPU cycles required for computing one input bit at user $i$. Then the total number of CPU cycles required for the $q_i$ bits is $C_iq_i$. By applying dynamic voltage and frequency scaling (DVFS) techniques\cite{Bar14,Mao17}, user $i$ can control the energy consumption for local task execution by adjusting the CPU frequency $f_{i,n}$ for each cycle $n$, where $f_{i,n}\in(0,f_i^{\rm max}]$, $n\in\{1,\ldots,C_iq_i\}$, and $f_i^{\rm max}$ denotes user $i$'s maximum CPU frequency.\footnote{Note that in practice, each CPU frequency $f_{i,n}$ can only be an integer chosen from a finite set. However, such an integer constraint may make the design problem a mixed-integer one that is NP-hard in general. To avoid this, we model $f_{i,n}$ as continuous variables to provide a performance upper-bound for the practical cases with discrete CPU frequencies.} With $f_{i,n}$'s, the execution time for local computing at user $i$ is $\sum_{n=1}^{C_iq_i} \frac{1}{f_{i,n}}$. As each user $i\in{\cal K}$ needs to accomplish the task execution within a block, the execution time cannot exceed the block length $T$, i.e.,
\begin{align}\label{eq.loc_latency}
\sum_{n=1}^{C_iq_i} \frac{1}{f_{i,n}} \leq T,~~\forall i\in{\cal K}.
\end{align}
 Under the assumption of a low CPU voltage that normally holds for low-power devices, the consumed energy for local computing at user $i\in{\cal K}$ could be expressed as \cite{Burd96}
\begin{align}\label{eq.energy_loc}
E_{{\rm loc},i} = \sum_{n=1}^{C_iq_i} \kappa_i f_{i,n}^2,
\end{align}
where $\kappa_i$ is the effective capacitance coefficient that depends on the chip architecture at user $i$.

\subsection{Energy Harvesting Constraints at Users}
As each user $i\in{\cal K}$ is powered by the WPT from the AP to achieve self-sustainable operation, the so-called energy harvesting constraint needs to be imposed such that the totally consumed energy at the user cannot exceed the harvested energy $E_i$ in (\ref{eq.energy_harvested}) per block. By combining the computation offloading energy in (\ref{eq.energy_offl}) and the local computation energy in (\ref{eq.energy_loc}), the total energy consumed by user $i$ within the block is $E_{\text{offl},i}+E_{\text{loc},i}$. Therefore, we must have per user $i\in{\cal K}$:\footnote{Note that in (\ref{eq.energy_harvest_const}) we consider that the totally consumed energy should not exceed the totally harvested one, instead of the ``energy causality'' in conventional energy harvesting communications (see, e.g., \cite{Li15}). This consideration implies that at the beginning of the block each user has sufficiently large energy storage, such that the stored energy will never be used up at any time within each block and the energy storage level will be refilled via energy harvesting by the end of each block.}
\begin{equation} \label{eq.energy_harvest_const}
E_{\text{loc},i}+E_{\text{offl},i} \leq E_i.
\end{equation}

\section{Computation Latency Constrained Energy Minimization}

\subsection{Problem Formulation}
Under the above setup, we pursue an energy-efficient MEC-WPT design by considering a computation latency constrained energy minimization problem. Suppose that each user $i\in{\cal K}$ has a computation task with $R_i>0$ input bits, which needs to be successfully executed before the end of the block. In this case, the sum of the number of offloaded bits $\ell_i$ and the number of locally computed bits $q_i$ should be equal to $R_i$, i.e., we have $q_i=R_i-\ell_i$,~$\forall i\in{\cal K}$.

We aim to minimize the energy consumption at the AP (including the energy consumption $\sum_{i=1}^K \alpha \ell_i$ in (\ref{eq.linear_energy_AP}) for computation and $T{\rm tr}({\bm Q})$ for WPT) while ensuring the successful execution of the $K$ users' computation tasks per time block. To this end, we jointly optimize the energy transmit covariance matrix $\boldsymbol{Q}$ at the AP, the local CPU frequencies $\{f_{i,1},\ldots,f_{i,C_i(R_i-\ell_i)}\}$, and the numbers of offloaded bits $\ell_i$'s at the users, as well as the time allocation $t_i$'s among different users. Let $\boldsymbol{t}\triangleq[t_1,\ldots,t_K]^\dagger$, $\boldsymbol{\ell}\triangleq[\ell_1,\ldots,\ell_K]^\dagger$, and $\boldsymbol{f}\triangleq [f_{1,1},\ldots,f_{K,C_K(R_K-\ell_K)}]^{\dagger}$. Mathematically, the latency-constrained energy minimization problem is formulated as
\begin{subequations}\label{eq.prob1}
\begin{align}
&({\cal P}1): ~ \min_{\boldsymbol{Q}\succeq \boldsymbol{0},\boldsymbol{t},\boldsymbol{\ell}, \boldsymbol{f}} ~~ T{\rm tr}(\boldsymbol{Q})+\sum_{i=1}^K \alpha\ell_i \\
 {\rm s.t.} & \sum_{n=1}^{C_i(R_i-\ell_i)} \frac{1}{f_{i,n}} \leq T,~~\forall i\in{\cal K} \\
 & \sum_{n=1}^{C_i(R_i-\ell_i)} \kappa_i f_{i,n}^2 +\frac{t_i}{\tilde{g}_i}\beta\left(\frac{\ell_i}{t_i}\right) +p_{c,i} t_i- T\zeta{\rm tr}(\boldsymbol{Q}\boldsymbol{H}_i) \leq 0,  \notag \\
 &~~~~~~~~~~~~~~~~~~~~~~~~~~~~~~~~~~~~~~~~~~~~~~\forall i\in{\cal K} \\
 & ~~\sum_{i=1}^K t_i \leq T,~~t_i\geq 0,~~0 \leq \ell_i \leq R_i,~~\forall i\in{\cal K} \\
 &~~0\leq f_{i,n}\leq f_i^{\rm max},~\forall n,~\forall i\in{\cal K}.
\end{align}
\end{subequations}
Here, the constraints in (\ref{eq.prob1}b) and (\ref{eq.prob1}c) represent the $K$ users' individual local computing latency and energy harvesting constraints, respectively. Note that due to the non-convex nature of (\ref{eq.prob1}b) and (\ref{eq.prob1}c), problem $({\cal P}1)$ is non-convex in the current form. However, we can transform it into a convex form and find the well-structured optimal solution, as will be shown in the next subsection.

\subsection{Optimal Solution to Problem $({\cal P}1)$}

In this subsection, we provide the optimal solution to the computation latency constrained energy minimization problem $({\cal P}1)$. To cope with the non-convex constraints in (\ref{eq.prob1}b) and (\ref{eq.prob1}c), we first establish the following lemma.

\begin{lemma}\label{lem.CPU}
Given the number of offloaded bits $\boldsymbol{\ell}$, the optimal solution of the local CPU frequencies $f_{i,n}$'s to problem $({\cal P}1)$ should satisfy
\begin{equation}\label{eq.opt_CPU}
f_{i,1}=\ldots=f_{i,C_i(R_i-\ell_i)}={C_i(R_i-\ell_i)}/{T},~~\forall i\in{\cal K}.
\end{equation}
\end{lemma}
\begin{IEEEproof}
See Appendix A.
\end{IEEEproof}

Lemma~\ref{lem.CPU} indicates that at each user $i\in{\cal K}$, the local CPU frequencies for different CPU cycles are identical at the optimality. Hence, problem $({\cal P}1)$ can be equivalently reformulated as
\begin{subequations}\label{eq.prob1_1}
\begin{align}
&({\cal P}1.1):~\min_{\boldsymbol{Q}\succeq \boldsymbol{0},\boldsymbol{t},\boldsymbol{\ell}} ~~ T{\rm tr}(\boldsymbol{Q}) +\sum_{i=1}^K \alpha \ell_i\\
 {\rm s.t.}&~~ \sum_{i=1}^K t_i \leq T  \\
 & ~~\frac{\kappa_iC^3_i(R_i-\ell_i)^3}{T^2}+\frac{t_i}{\tilde{g}_i}\beta\left(\frac{\ell_i}{t_i}\right)+p_{c,i}t_i -  T \zeta {\rm tr}(\boldsymbol{Q}\boldsymbol{H}_i) \leq 0\notag\\
 &~~~~~~~~~~~~~~~~~~~~~~~~~~~~~~~~~~~~~~~~~~~~~~~~\forall i\in{\cal K} \\
 & ~~0 \leq \ell_i \leq R_i,~~ t_i\geq 0, ~~\forall i\in{\cal K}.
\end{align}
\end{subequations}
As $\beta(x)$ is convex as a function of $x\geq 0$, its perspective function $\frac{t_i}{\tilde{g}_i}\beta\left(\frac{\ell_i}{t_i}\right)$ is jointly convex with respect to $t_i\geq 0$ and $\ell_i\geq 0$\cite{BoydBook}. As a result, the energy harvesting constraints in (\ref{eq.prob1_1}c) become convex. Furthermore, since the objective function in (\ref{eq.prob1_1}a) is affine and the other constraints are all convex, problem $({\cal P}1.1)$ is convex and can thus be optimally solved by standard convex optimization techniques. Nevertheless, to gain engineering insights, we derive its optimal solution in a semi-closed form by leveraging the Lagrange duality method \cite{BoydBook}.

Let $\mu\geq 0$ and $\lambda_i\geq 0$ denote the dual variables associated with the time-allocation constraint in (\ref{eq.prob1_1}b) and the $i$-th energy harvesting constraint in (\ref{eq.prob1_1}c), $\forall i\in{\cal K}$, respectively. Then the partial Lagrangian of $({\cal P}1.1)$ is expressed as
\begin{align}\label{eq.par_L_1}
 {\cal L}_1\left(\boldsymbol{Q},\boldsymbol{t},\boldsymbol{\ell},\boldsymbol{\lambda},\mu\right)  = & T{\rm tr}\left( \left( {\bm I}-\sum_{i=1}^K\zeta\lambda_i{\bm H}_i\right)\boldsymbol{Q} \right) -\mu T \notag \\
 &+ \sum_{i=1}^K \left( \alpha \ell_i + \frac{\lambda_i\kappa_iC^3_i(R_i-\ell_i)^3}{T^2}\right.\notag \\
 & \left. +\frac{\lambda_it_i}{\tilde{g}_i}\beta\left(\frac{\ell_i}{t_i}\right)
 +\lambda_ip_{c,i}t_i + \mu t_i \right),
\end{align}
where $\boldsymbol{\lambda}\triangleq [\lambda_1,\ldots,\lambda_K]^{\dagger}$. Accordingly, the dual function is given by
\begin{align} \label{eq.dual_obj1}
\Phi(\boldsymbol{\lambda},\mu) =& \min_{\boldsymbol{Q}\succeq \boldsymbol{0},\;\boldsymbol{t},\;\boldsymbol{\ell}} ~~{\cal L}\left(\boldsymbol{Q},\boldsymbol{t},\boldsymbol{\ell},\boldsymbol{\lambda},\mu\right) \\
~~{\rm s.t.}& ~~0\leq \ell_i \leq R_i,~~t_i\geq 0,~~\forall i \in{\cal K}. \notag
\end{align}
Consequently, the dual problem of $({\cal P}1.1)$ is
\begin{subequations} \label{eq.dual_prob1}
\begin{align}
 ({\cal D}1.1): ~~&\max_{\boldsymbol{\lambda},\;\mu} ~~\Phi(\boldsymbol{\lambda},\mu) \\
 {\rm s.t.} ~~
 &\boldsymbol{F}(\boldsymbol{\lambda}) \triangleq \boldsymbol{I}-\sum_{i=1}^K \zeta\lambda_i \boldsymbol{H}_i\succeq \boldsymbol{0} \\
& \mu \geq 0,~~\lambda_i\geq 0 ,~~\forall i\in {\cal K}.
\end{align}
\end{subequations}
Note that the constraint $\boldsymbol{F}(\boldsymbol{\lambda}) \succeq \boldsymbol{0}$ is necessary to ensure the dual function $\Phi({\bm \lambda},\mu)$ to be bounded from below (as proved in Appendix~B). We denote the feasible set of $(\boldsymbol{\lambda},\mu)$ characterized by (\ref{eq.dual_prob1}b) and (\ref{eq.dual_prob1}c) as ${\cal S}$.

Since problem $({\cal P}1.1)$ is convex and satisfies the Slater's condition, strong duality holds between $({\cal P}1.1)$ and its dual problem $({\cal D}1.1)$\cite{BoydBook}. As a result, we can solve $({\cal P}1.1)$ by equivalently solving $({\cal D}1.1)$. In the following, we first obtain the dual function $\Phi(\boldsymbol{\lambda},\mu)$ for any given $(\boldsymbol{\lambda},\mu)\in{\cal S}$, and then find the optimal dual variables ${\bm \lambda}$ and $\mu$ to maximize $\Phi(\boldsymbol{\lambda},\mu)$ using the ellipsoid method\cite{Boyd_Ellipsoid}. For convenience of presentation, let $(\boldsymbol{Q}^*, \boldsymbol{t}^*,\boldsymbol{\ell}^*)$ denote the optimal solution to problem (\ref{eq.dual_obj1}) for given $\boldsymbol{\lambda}$ and $\mu$, $(\boldsymbol{Q}^{\rm{opt}}, \boldsymbol{t}^{\rm{opt}},\boldsymbol{\ell}^{\rm{opt}})$ denote the optimal primary solution to $({\cal P}1.1)$, and $(\boldsymbol{\lambda}^{\rm{opt}},\mu^{\rm{opt}})$ denote the optimal dual solution to $({\cal D}1.1)$.

\subsubsection{Evaluating the Dual Function $\Phi(\boldsymbol{\lambda},\mu)$}

First, we obtain the dual function $\Phi({\bm \lambda},\mu)$ in (\ref{eq.dual_obj1}) for any given $(\boldsymbol{\lambda},\mu)\in{\cal S}$. To this end, problem (\ref{eq.dual_obj1}) can be decomposed into $(K+1)$ subproblems as follows, one for optimizing $\boldsymbol{Q}$ and the other $K$ for jointly optimizing $t_i$'s and $\ell_i$'s.
\begin{align}\label{eq.Q1_iter}
\min_{\boldsymbol{Q}}~~{\rm tr}\left(\boldsymbol{Q} \boldsymbol{F}(\boldsymbol{\lambda}) \right) ~~~~{\rm s.t.} ~~ \boldsymbol{Q}\succeq \boldsymbol{0}.
\end{align}
\vspace{-0.5cm}
\begin{subequations}\label{eq.La1_dist}
\begin{align}
\min_{t_i,\ell_i}&~~
\alpha \ell_i+ \frac{\lambda_i \kappa_i C^3_i(R_i-\ell_i)^3}{T^2}+\frac{\lambda_it_i}{\tilde{g}_i}\beta\left(\frac{\ell_i}{t_i}\right)+\lambda_i p_{c,i}t_i+\mu t_i  \\
~~{\rm s.t.}& ~~0\leq \ell_i \leq R_i, ~~t_i\geq 0,
\end{align}
\end{subequations}
where each subproblem $i$ in (\ref{eq.La1_dist}) is for the user $i\in{\cal K}$. Under the condition of $\boldsymbol{F}(\boldsymbol{\lambda})\succeq \boldsymbol{0}$, it is evident that the optimal value of problem (\ref{eq.Q1_iter}) is zero and its optimal solution $\boldsymbol{Q}^*$ can be any positive semidefinite matrix in the null space of $\boldsymbol{F}(\boldsymbol{\lambda})$. Without loss of optimality, we simply set $\boldsymbol{Q}^*=\boldsymbol{0}$ for the purpose of obtaining the dual function $\Phi(\boldsymbol{\lambda},\mu)$ and computing the optimal dual solution.\footnote{Note that $\boldsymbol{Q}^*=\boldsymbol{0}$ is not a unique optimal solution to problem (\ref{eq.Q1_iter}) when $\boldsymbol{F}(\boldsymbol{\lambda})$ is rank-deficient, i.e., ${\rm rank}(\boldsymbol{F}(\boldsymbol{\lambda}))<N$.} Note that ${\bm Q}^*={\bm 0}$ is generally not the optimal primary solution to $({\cal P}1.1)$. As a result, after finding the optimal dual solution $({\bm \lambda}^{\rm{opt}},\mu^{\rm{opt}})$, we need to use an additional step to retrieve the optimal primary solution of $\boldsymbol{Q}^{\rm{opt}}$ to $({\cal P}1.1)$, as will be shown in Section IV-C.

For the $i$-th subproblem in (\ref{eq.La1_dist}), it is convex and satisfies the Slater's condition. Based on the Karush-Kuhn-Tucker (KKT) conditions\cite{BoydBook}, one can obtain the optimal solution $(t_i^*,\ell_i^*)$ to (\ref{eq.La1_dist}) in a semi-closed form, as stated in the following lemma.

\begin{lemma}\label{lem.t_l_sol1}
For any given $({\bm \lambda},\mu)\in{\cal S}$, the optimal solution $(t^*,\ell_i^*)$ to problem (\ref{eq.La1_dist}) can be obtained as follows.
 \begin{itemize}
 \item If $\lambda_i = 0$, we have $\ell^*_i = 0$ and $t^*_i=0$;
 \item If $\lambda_i > 0$, we have
\begin{align}\label{eq.prim1_sol}
\ell^*_i &= \left[ R_i- \sqrt{\frac{T^2}{3\kappa_iC^3_i} \left( \frac{\alpha}{\lambda_i} +\frac{\sigma^2 \ln 2}{B\tilde{g}_i}2^{\frac{r^*_i}{B}} \right) } \right]^+ \\
 t^*_i & =   {\ell^*_i}/{r_i^*},
\end{align}
where $r_i^* \triangleq \frac{B}{\ln 2} \left(W_0\left(\frac{\tilde{g}_i}{\sigma^2 e}\left( \frac{\mu}{\lambda_i}+p_{{\rm c},i} \right)-\frac{1}{e}\right)+1\right)$ denotes the offloading rate of user $i$, $W_0(x)$ is the principal branch of the Lambert $W$ function defined as the solution for $W_0(x)e^{W_0(x)}=x$\cite{LambertW}, and $e$ is the base of the natural logarithm.
\end{itemize}

\end{lemma}
\begin{IEEEproof}
See Appendix C.
\end{IEEEproof}

By combining Lemma~\ref{lem.t_l_sol1} and $\boldsymbol{Q}^*=\boldsymbol{0}$, the dual function $\Phi(\boldsymbol{\lambda},\mu)$ can be evaluated for any given $(\boldsymbol{\lambda},\mu)\in{\cal S}$.

\subsubsection{Obtaining the Optimal $\boldsymbol{\lambda}^{\rm{opt}}$ and $\mu^{\rm{opt}}$ to Maximize $\Phi(\boldsymbol{\lambda},\mu)$}\label{subsubsection:dual_opt}

Having obtained $({\bm Q}^*,{\bm \ell}^*,{\bm t}^*)$ for given ${\bm \lambda}$ and $\mu$, we can next solve the dual problem $({\cal D}1.1)$ to maximize $\Phi({\bm \lambda},\mu)$. Note that the dual function $\Phi(\boldsymbol{\lambda},\mu)$ is concave but non-differentiable in general\cite{BoydBook}. Hence, we use subgradient based methods, e.g., the ellipsoid method \cite{Boyd_Ellipsoid}, to obtain the optimal $\boldsymbol{\lambda}^{\rm{opt}}$ and $\mu^{\rm{opt}}$ for problem $({\cal D}1.1)$. The basic idea of the ellipsoid method is to find a series of ellipsoids to localize the optimal dual solution $\boldsymbol{\lambda}^{\rm{opt}}$ and $\mu^{\rm{opt}}$\cite{Boyd_Ellipsoid}. To start with, we choose a given $(\boldsymbol{\lambda},\mu)\in{\cal S}$ as the center of the initial ellipsoid and set its volume to be sufficiently large to contain $(\boldsymbol{\lambda}^{\rm{opt}},\mu^{\rm{opt}})$. Then, at each iteration, we update the dual variables $({\bm \lambda},\mu)$ based on the subgradients of both the objective function and the constraint functions, and accordingly construct a new ellipsoid with reduced volume. When the volume of the ellipsoid is reduced below a certain threshold, the iteration will terminate and the center of the ellipsoid is chosen to be the optimal dual solution $(\boldsymbol{\lambda}^{\rm{opt}},\mu^{\rm{opt}})$. More details can be referred to in~\cite{Boyd_Ellipsoid}.

To implement the ellipsoid method, it remains to determine the subgradients of both the objective function in (\ref{eq.dual_prob1}a) and the constraint functions in (\ref{eq.dual_prob1}b) and (\ref{eq.dual_prob1}c). For the objective function $\Phi({\bm \lambda},\mu)$ in (\ref{eq.dual_prob1}a), one subgradient is given by \cite{Boyd_Ellipsoid}
\begin{subequations}\label{eq.obj1_subg}
\begin{align}
&\left[\frac{\kappa_1C_1^3(R_1-\ell^*_1)^3}{T^2}+\frac{t^*_1}{\tilde{g}_1}\beta\left(\frac{\ell^*_1}{t^*_1}\right)+p_{{\rm c},1}t^*_1,\ldots,\right. \notag \\
&~~\left. \frac{\kappa_KC_K^3(R_K-\ell^*_K)^3}{T^2}+\frac{t^*_K}{\tilde{g}_K}\beta\left(\frac{\ell^*_K}{t^*_K}\right)+p_{{\rm c},K}t^*_K, \sum_{i=1}^K t^*_i -T\right]^\dagger.
\end{align}
\end{subequations}

As for the constraint $\boldsymbol{F}(\boldsymbol{\lambda})\succeq \boldsymbol{0}$ in (\ref{eq.dual_prob1}b), we have the following lemma.

\begin{lemma} \label{lem.F_sub}
Let $\boldsymbol{v}\in \mathbb{C}^{N\times 1}$ be the eigenvector corresponding to the smallest eigenvalue of $\boldsymbol{F}(\boldsymbol{\lambda})$, i.e., $\boldsymbol{v}=\arg\min_{\|\boldsymbol{\xi}\|=1} \boldsymbol{\xi}^H\boldsymbol{F}(\boldsymbol{\lambda})\boldsymbol{\xi}$. Then the constraint $\boldsymbol{F}(\boldsymbol{\lambda})\succeq \boldsymbol{0}$ is equivalent to the constraint of $ \boldsymbol{v}^H\boldsymbol{F}(\boldsymbol{\lambda})\boldsymbol{v} \geq 0$, and the subgradient of $ \boldsymbol{v}^H\boldsymbol{F}(\boldsymbol{\lambda})\boldsymbol{v}$ at the given $\boldsymbol{\lambda}$ and $\mu$ is
\begin{equation} \label{eq.subg_lem}
\left[ \zeta\boldsymbol{v}^H\boldsymbol{H}_1\boldsymbol{v},\ldots,\zeta\boldsymbol{v}^H\boldsymbol{H}_K\boldsymbol{v}, 0 \right]^{\dagger}.
\end{equation}
\end{lemma}

\begin{IEEEproof}
See Appendix D.
\end{IEEEproof}

Furthermore, the subgradient of $\lambda_i\geq 0$ in (\ref{eq.dual_prob1}c) is given by the elementary vector $\boldsymbol{e}_i\in {\mathbb{R}^{(K+1)\times1}}$ (i.e., $\boldsymbol{e}_i$ is of all zero entries except for the $i$-th entry being one), $\forall i\in{\cal K}$, while that of $\mu\geq 0$ is $\boldsymbol{e}_{K+1}$. By using this together with (\ref{eq.obj1_subg}) and (\ref{eq.subg_lem}), the ellipsoid method can be applied to efficiently update $\boldsymbol{\lambda}$ and $\mu$ towards the optimal $\boldsymbol{\lambda}^{\rm{opt}}$ and $\mu^{\rm{opt}}$ for $({\cal D}1.1)$.

\subsubsection{Finding the Optimal Primary Solution to $({\cal P}1)$}

With $\boldsymbol{\lambda}^{\rm{opt}}$ and $\mu^{\rm{opt}}$ obtained, it remains to determine the optimal primary solution to $({\cal P}1.1)$ (or equivalently $({\cal P}1)$). Specifically, by replacing $\boldsymbol{\lambda}$ and $\mu$ with $\boldsymbol{\lambda}^{\rm{opt}}$ and $\mu^{\rm{opt}}$ in Lemma~\ref{lem.t_l_sol1}, one can obtain the optimal $(\boldsymbol{t}^{\rm{opt}},\boldsymbol{\ell}^{\rm{opt}})$ for $({\cal P}1)$ in a semi-closed form. Furthermore, by substituting $\boldsymbol{\ell}^{\rm{opt}}$ in Lemma~\ref{lem.CPU}, one can then obtain the optimal local CPU frequencies $\{f^{\rm{opt}}_{i,n}\}$ for the $K$ users. However, we cannot directly obtain the optimal energy transmit covariance matrix ${\bm Q}^{\rm{opt}}$ for $({\cal P}1)$ from the solution to problem (\ref{eq.Q1_iter}), since its solution is non-unique in general. Therefore, we adopt an additional step to obtain ${\bm Q}^{\rm{opt}}$ by solving a semidefinite program (SDP), which corresponds to solving problem $({\cal P}1.1)$ for ${\bm Q}$ under the given $({\bm t}^{\rm{opt}},{\bm \ell}^{\rm{opt}})$.

We can then readily establish the following proposition.

\begin{proposition}\label{prop:prop}
The optimal solution $(\{f^{\rm{opt}}_{i,n}\},\boldsymbol{Q}^{\rm{opt}},\boldsymbol{t}^{\rm{opt}},\boldsymbol{\ell}^{\rm{opt}})$ for problem $({\cal P}1)$ is given by
\begin{align}
& \ell_i^{\rm{opt}}  =
\begin{cases}
\left[ R_i- \sqrt{\frac{T^2}{3\kappa_iC_i^3} \left( \frac{\alpha}{\lambda^{\rm{opt}}_i} +\frac{\sigma^2 \ln 2}{B\tilde{g}_i}2^{\frac{r^{\rm{opt}}_i}{B}} \right) } \right]^+, &{\rm if}~\lambda_i^{\rm{opt}} >  0, \\
0,&{\rm if}~\lambda_i^{\rm{opt}} = 0,\\
\end{cases}  \label{eq.opt_l} \\
& t_i^{\rm{opt}}  =
\begin{cases}
\ell_i^{\rm{opt}}/r_i^{\rm{opt}}, &{\rm if}~\lambda_i^{\rm{opt}} > 0, \\
0,&{\rm if}~\lambda_i^{\rm{opt}} = 0,\\
\end{cases} \label{eq.opt_t} \\
&f^{\rm{opt}}_{i,1} = \ldots = f^{\rm{opt}}_{i,C_i(R_i-\ell_i^{\rm{opt}})} = {C_i(R_i-\ell^{\rm opt}_i)}/{T}, ~~\forall i\in{\cal K},\label{eq.primary_CPU}
\end{align}
and
\begin{align}\label{eq.opt_Q}
\boldsymbol{Q}^{\rm{opt}} &= \arg \min_{\boldsymbol{Q}\succeq \boldsymbol{0}} ~~ T{\rm tr}(\boldsymbol{Q}) \notag \\
\rm{s.t.}  & ~~ \frac{\kappa_iC_i^3(R_i-\ell_i^{\rm{opt}})^3}{T^2}+\frac{t_i^{\rm{opt}}}{\tilde{g}_i}\beta\left( r_i^{\rm{opt}} \right)+p_{c,i}t^{\rm{opt}}_i \notag \\
&~~~~~~~~~~~~~~~~~~~~-T \zeta {\rm tr}(\boldsymbol{Q}\boldsymbol{H}_i) \leq 0,~~\forall i\in{\cal K},
\end{align}
\end{proposition}
where
\begin{equation} \label{eq.opt_r}
r^{\rm{opt}}_i \triangleq \frac{B}{\ln 2} \left(W_0\left(\frac{\tilde{g}_i}{\sigma^2 e}\left( \frac{\mu^{\rm{opt}}}{\lambda^{\rm{opt}}_i}+p_{c,i} \right)-\frac{1}{e}\right)+1\right)
\end{equation}
corresponds to the optimal offloading rate for user $i$, $\forall i\in{\cal K}$.

Proposition \ref{prop:prop} can be verified by simply combining Lemmas~\ref{lem.CPU} and~\ref{lem.t_l_sol1}; hence, we omit its detailed proof for conciseness. Note that (\ref{eq.opt_Q}) is an instance of SDP, which can thus be efficiently solved by off-the-shelf solvers, e.g., CVX \cite{cvx}.

Summarizing, we present Algorithm~\ref{alg1} to solve the computation latency constrained energy minimization problem $({\cal P}1)$.

\begin{algorithm}
\caption{\it for Solving the Energy Minimization Problem $({\cal P}1)$}\label{alg1}
\begin{algorithmic}[1]
\State
{\bf Initialization:}
Given an ellipsoid ${\cal E}((\boldsymbol{\lambda},\mu),\boldsymbol{A})$ containing $(\boldsymbol{\lambda}^{\rm{opt}},\mu^{\rm{opt}})$, where $(\boldsymbol{\lambda},\mu)$ is the center of ${\cal E}$ and $\boldsymbol{A}\succ \boldsymbol{0}$ characterizes the volume of ${\cal E}$.
\State
{\bf Repeat:}
\begin{itemize}
\item For each user $i\in{\cal K}$, obtain $(t^*_i,\ell^*_i)$ by Lemma~\ref{lem.t_l_sol1} under given $\lambda_i$ and $\mu$;
\item Compute the subgradients of the objective function and the constraints of $({\cal D}1.1)$ as in Section III-B.2;
\item Update $\boldsymbol{\lambda}$ and $\mu$ using the ellipsoid method \cite{Boyd_Ellipsoid};
\end{itemize}
  \State
 {\bf Until} $\boldsymbol{\lambda}$ and $\mu$ converge within a prescribed accuracy.
\State
 {\bf Set} $(\boldsymbol{\lambda}^{\rm{opt}},\mu^{\rm{opt}}) \gets (\boldsymbol{\lambda},\mu)$.
 \State
{\bf Output}: Obtain $(\boldsymbol{t}^{\rm{opt}},\boldsymbol{\ell}^{{\rm{opt}}})$, $\{f^{\rm{opt}}_{i,n}\}$, and compute $\boldsymbol{Q}^{\rm{opt}}$ by (\ref{eq.opt_Q}).
\end{algorithmic}
\end{algorithm}

\begin{remark}\label{remark1}
Proposition \ref{prop:prop} shows that the optimal joint computing and offloading design has the following interesting properties to minimize the energy consumption at the AP.
\begin{enumerate}
\item
First, if the energy harvesting constraint is not tight for user $i$ (i.e., user $i$ harvests sufficient wireless energy), then no computation offloading is required and user $i$ should compute all the tasks locally (i.e., $\ell_i^{\rm opt}=0$). This can be explained based on the complementary slackness condition\cite{BoydBook}, i.e.,
\begin{align}\label{eq:lambda_opt}
&\lambda_i^{\rm opt} \left(\frac{\kappa_iC_i^3(R_i-\ell_i^{\rm opt})^3}{T^2} +\frac{t^{\rm opt}_i}{\tilde{g}_i}\beta\left(r_i^{\rm opt}\right) +p_{c,i} t^{\rm opt}_i \right. \notag \\
&~~~~~~~~~~~~~~ -T\zeta{\rm tr}\left(\boldsymbol{Q}^{\rm opt}\boldsymbol{H}_i\right) \Bigg) =0,~~\forall i\in{\cal K}.
\end{align}
In this case, if the energy harvesting constraint is not tight for user $i$, then based on \eqref{eq:lambda_opt} we have $\lambda^{\rm opt}_i=0$, and accordingly $\ell_i^{\rm opt}=0$ holds from \eqref{eq.opt_l}. This property is intuitive: when the user has sufficient energy to accomplish the tasks locally, there is no need to employ computation offloading that incurs additional energy consumption for the MEC server's computation at the AP.

\item
Next, it is always beneficial to leave some bits for local computing at each user $i\in{\cal K}$, i.e., $\ell_i^{\rm opt}<R_i$ always holds (see (\ref{eq.opt_l})). In other words, offloading all the bits to the AP is always suboptimal. This is because when $\ell_i^{\rm opt}\to R_i$, the marginal energy consumption of local computing is almost zero, and thus it is beneficial to leave some bits for local computing in this case.

 \item
Furthermore, it is observed that for each user $i$, more stringent the energy harvesting constraint is (or the associated dual variable $\lambda_i^{\rm opt}$ is larger), more bits should be offloaded to the AP with a smaller offloading rate $r_i^{\rm opt}$. This property follows based on \eqref{eq.opt_l} and \eqref{eq.opt_r}, in which a larger $\lambda_i^{\rm opt}$ admits a larger $\ell_i^{\rm opt}$ and a smaller $r_i^{\rm opt}$.

\item
Finally, the number of offloaded bits $\ell^{\rm opt}_i$ and the offloading rate $r^{\rm opt}_i$ for each user $i$ are affected by the channel gain $\tilde{g}_i$, the block length $T$, the circuit power $p_{{\rm c},i}$, and the MEC energy consumption $\alpha$ per offloaded bit in the following way: 1) when the channel condition becomes better (i.e., $\tilde{g}_i$ becomes larger), both $\ell^{\rm opt}_i$ and $r^{\rm opt}_i$ increase, and thus user $i$ is likely to offload more bits with a higher offloading rate; 2) a higher circuit power $p_{c,i}$ at the user leads to a higher offloading rate $r^{\rm opt}_i$; 3) when $T$ or $\alpha$ increases, $\ell^{\rm opt}_i$ reduces and thus fewer bits are offloaded to the AP.
\end{enumerate}
\end{remark}

\section{Numerical Results}

In this section, numerical results are provided to gauge the performance of the proposed design with joint WPT, offloading, and computing optimization, as compared to the following four benchmark schemes.
\begin{enumerate}
\item {\em Local computing only:} each user $i\in{\cal K}$ accomplishes its computation task by only local computing. This scheme corresponds to solving problem $({\cal P}1)$ by setting $\ell_i=0$, $\forall i\in{\cal K}$.

\item {\em Computation offloading only:} each user $i\in{\cal K}$ accomplishes its computation task by fully offloading the computation bits to the AP. This scheme corresponds to solving $({\cal P}1)$ by setting $f_{i,n}=0$, $\forall n$, $\forall i\in{\cal K}$, as well as $\ell_i=R_i$ for $({\cal P}1)$, $\forall i\in{\cal K}$.

\item {\em Joint design with isotropic WPT:} the $N$-antenna AP radiates the RF energy isotropically or omni-directionally over all directions by setting ${\bm Q}=p{\bm I}$, where $p\geq 0$ denotes the transmit power at each antenna. This scheme corresponds to solving problem $({\cal P}1)$ by replacing $\boldsymbol{Q}$ as $p\boldsymbol{I}$ with $p$ being another optimization variable.

\item {\em Separate MEC-WPT design:} this scheme separately designs the computation offloading for MEC and the energy beamforming for WPT \cite{Huang16,Zeng17}. First, the $K$ users minimize their sum-energy consumption subject to the users' individual computation latency constraints \cite{Huang16}. Then, under the constraints of energy demand at the $K$ users, the AP designs the transmit energy beamforming with minimum energy consumption \cite{Zeng17}.
\end{enumerate}

In the simulations, the EH efficiency is set as $\zeta=0.3$. The system parameters are set as (unless stated otherwise): the number of the AP antennas $N=4$, $C_i =10^3$ cycles/bit, $\kappa_i=10^{-28}, \forall i\in\mathcal K$ \cite{Huang16}, the circuit power $p_{c,i}=10^{-4}$ Watt (W), the energy consumption per offloaded bit by the MEC server $\alpha=10^{-4}$ Joule/bit, the receiver noise power $\sigma^2 = 10^{-9}$~W, and the spectrum bandwidth for offloading $B=2$ MHz. By considering a Rayleigh fading channel model, the wireless channel from the AP to each user $i\in\mathcal K$ is set as
\begin{align}
\mv h_i = \theta_0 d_i^{-3} \mv {\bar h}_i,~~\mv g_i = \theta_0 d_i^{-3} \mv {\bar g}_i,
\end{align}
where $\mv {\bar h}_i\sim {\cal CN}(\boldsymbol{0},\mv I)$ and $\mv {\bar g}_i\sim {\cal CN}(\boldsymbol{0},\mv I)$, $\forall i\in\mathcal K$, is an independent and identically distributed (i.i.d.) circularly symmetric complex Gaussian (CSCG) random vector with zero mean and covariance $\mv I$; $\theta_0 = 6.25\times 10^{-4}$ (i.e., $-32$ dB) corresponds to the channel power gain at a reference distance of one meter; $d_i$ denotes the distance from the AP to user $i\in{\cal K}$; and the path-loss exponent is assumed to be~3. The numerical results are obtained by averaging over 500 randomized channel realizations. Note that the simulation parameters are specifically chosen, but our approaches can be also applied to other system setups.

\subsection{Case with Homogeneous Users}
\begin{figure}
  \centering
  \includegraphics[width = 3.5in]{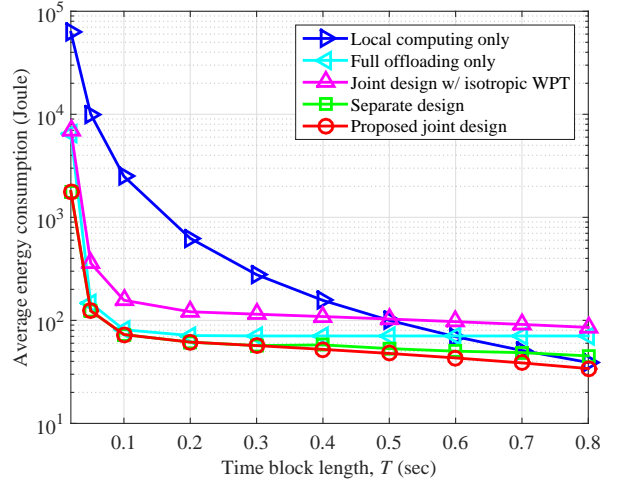}
  \caption{The average energy consumption at the AP versus the block length $T$.} \label{fig.perf_vs_T}
\end{figure}

\begin{figure}
  \centering
  \includegraphics[width = 3.5in]{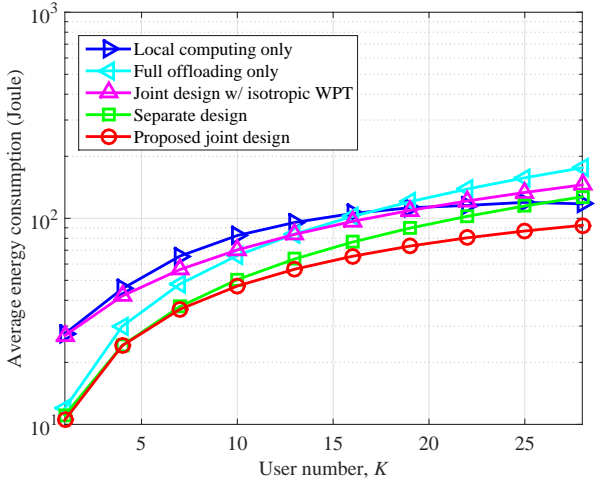}
  \caption{The average energy consumption at the AP versus the user number $K$.} \label{fig.perf_vs_K}
\end{figure}

\begin{figure}
  \centering
  \includegraphics[width = 3.5in]{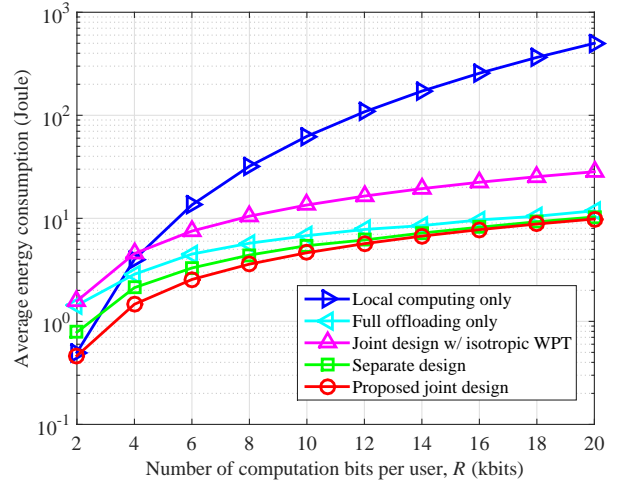}
  \caption{The average energy consumption at the AP versus the number of computation bits $R$ at each user.} \label{fig.perf_vs_R}
\end{figure}

\begin{figure}
  \centering
  \includegraphics[width = 3.5in]{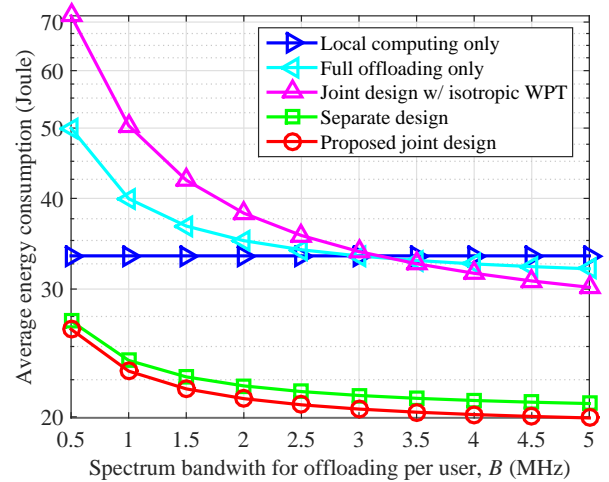}
  \caption{The average energy consumption at the AP versus the spectrum bandwidth $B$ for offloading.} \label{fig.perf_vs_BW}
\end{figure}

First, we consider the case with homogeneous users, where the distances from the AP to all the users are identical with $d_i=5$ meters, $\forall i\in{\cal K}$. The corresponding average power loss is set to be $5\times 10^{-6}$ (i.e., $-53$~dB). Additionally, the numbers of computation bits at all users are set to be identical, i.e., $R=R_i$, $\forall i\in{\cal K}$. Figs.~\ref{fig.perf_vs_T}--\ref{fig.perf_vs_BW} show the average energy consumption at the AP under different system parameters. It is observed that the proposed joint design achieves the lowest average energy consumption at the AP among all the five schemes. The joint design with isotropic WPT achieves a suboptimal performance due to the loss of multi-antenna energy beamforming gain. The suboptimal performance of the separate-design scheme implies the necessity of unified demand-supply optimization in wireless powered MEC systems.

Fig.~\ref{fig.perf_vs_T} shows the average energy consumption at the AP versus the time block length $T$, where $R=10$ kbits and $K=10$. First, with a small value of $T$ (e.g., $T=0.05$ sec), the benchmark schemes but the local-computing-only scheme are observed to achieve a near optimal performance close to that with the proposed joint design, while as $T$ increases, the energy consumption with the local-computing-only scheme significantly decreases, approaching that with the proposed joint design. It is also observed that the energy consumption with the full-offloading-only scheme remains almost unchanged when $T\geq 0.1$ sec. This is due to the fact that in this case, the optimal offloading time for all users is fixed to be around $0.1$~sec for saving the circuit energy consumption; hence, increasing $T$ cannot further improve the energy efficiency in this case. By contrast, the energy consumption with the local-computing-only scheme decreases monotonically as $T$ increases. This is because as $T$ increases, one can always lower down the CPU frequency to save energy for local computing. Finally, it is seen in Fig.~\ref{fig.perf_vs_T} both the separate-design and the equal-offloading-time-allocation schemes achieve a very similar performance in the interested time block regime.

Fig.~\ref{fig.perf_vs_K} depicts the average energy consumption versus the user number $K$, where $R=10$ kbits and $T=0.5$~sec. It is shown that the gain achieved by the proposed joint design becomes more significant as the user number $K$ becomes large. The full-loading-only scheme outperforms the local-computing-only scheme, but with a decreasing gain as $K$ increases. This is because in the full-offloading-only scheme, all users share the finite time block and the offloading energy consumption would increase drastically when $K$ becomes large. It is also observed that the performance of the equal-offloading-time-allocation scheme becomes closer to that of the proposed joint design with larger $K\geq 15$. This indicates that an equal offloading time is desirable for a large number of the users in order to minimize the energy consumption at the AP.

Fig.~\ref{fig.perf_vs_R} shows the average energy consumption at the AP versus the number of computation bits $R$ at each user, where $K=2$ and $T=0.05$~sec. It is shown that the average energy consumption by all the six schemes increases as $R$ becomes large, and the full-offloading-only scheme outperforms the local-computing-only one, especially when $R$ becomes large. This indicates that with large $R$ values, it is desirable to offload more computation bits to the AP in order to reduce the energy consumption. Furthermore, the full-offloading-only scheme is observed to achieve a near optimal performance close to that with the proposed joint design when $R$ becomes large. This is because the energy consumption per bit for offloading is significantly smaller than that for local computing in the large $R$ case. It is also observed that the separate-design scheme outperforms all the other benchmark schemes in this setup.

Fig.~\ref{fig.perf_vs_BW} shows the average energy consumption versus the spectrum bandwidth $B$ for offloading, where $K=6$, $T=0.5$~sec, and $R=50$~kbits. As expected, Fig.~\ref{fig.perf_vs_BW} shows that the energy consumption by the four schemes with offloading decreases as $B$ increases, and the one by the local-computing-only scheme remains unchanged. This indicates that a large value of $B$ not only implies a high offloading rate, but also helps save the energy consumption in computation offloading. It is also observed that at small $B$ values (e.g., $B\leq 3$ MHz), the local-computing-only scheme outperforms the full-offloading-only scheme, but it does not hold for large $B$ cases. This indicates that offloading becomes a better option than local-computing as $B$ increases.

\subsection{Case with Heterogeneous Users}

Next, we evaluate the performance of the wireless powered MEC system in the case with heterogeneous users. For the purpose of illustration, we focus on the scenario with only $K=2$ users. It is assumed that the distances from the AP to the two users (namely near and far users) are $d_1=2$ meters and $d_2\geq2$ meters, respectively. The time block length is set as $T=0.2$~sec.

\begin{figure}
  \centering
  \includegraphics[width = 3.5in]{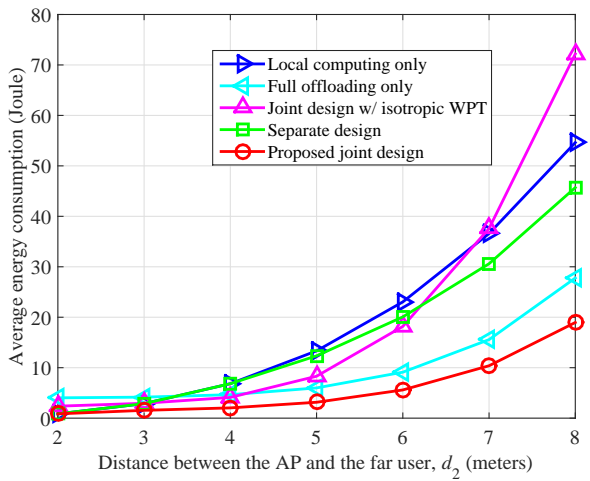}
  \caption{The average energy consumption at the AP versus the distance $d_2$ from the AP to the far user.} \label{fig.perf_vs_d}
\end{figure}

\begin{table*} [ht]%[htdp]
\caption{Offloaded Bits and Residual Energy at Users for the Proposed Optimal Joint Design Under Different $d_2$.} \label{tab.vs_d}
\centering
\begin{tabular}{c c c c c c c c}
\hline \hline
$d_2$ (meters) & 2 & 3 & 4 & 5 & 6 & 7 & 8 \\ [0.5ex]
\hline
Near user's offloaded bits $\ell^{\rm opt}_1$ (kbits) &  0.165 & 0.142 & 0.124 & 0.092 & 0.068 & 0.028 & 0.006 \\
Far user's offloaded bits $\ell^{\rm opt}_2$ (kbits) & 1.798 & 6.974 & 11.817 & 13.682 & 13.585 & 12.972 & 12.162 \\
%near user's offloaded time $t_1$ ($\times 10^{-2}$ sec) & 0.27 & 0.59 & 0.77 & 1.86 & 5.0 & 8.28 & 10.29\\
%far user's offloaded time $t_2$ ($\times10^{-2}$ sec) &  0.09 & 0.1 & 0.33 & 0.53 & 0.72 & 0.84 & 0.93\\
Near user's residual energy ($\times 10^{-5}$ Joule)& 0.007 & 0.026 & 0.062 & 0.531 & 3.276 & 9.218 & 21.105\\
 Far user's residual energy ($\times 10^{-5}$ Joule) & 0.003 & 0 & 0 & 0 & 0 & 0 & 0\\
[1ex]
 \hline
\end{tabular}
\end{table*}

\begin{table*} [ht]%[htdp]
\caption{Offloaded Bits and Residual Energy at Users for the Proposed Optimal Joint Design Under Different $R_2$.} \label{tab.vs_size}
\centering
\begin{tabular}{c c c c c  }
\hline \hline
$R_2$ (kbits) & 10 & 20 & 30 & 40 \\ [0.5ex]
\hline
Near user's offloaded bits $\ell^{\rm opt}_1$ (kbits) & 0.825 & 0.432 & 0.239 & 0.282  \\
Far user's offloaded bits $\ell_2^{\rm opt}$ (kbits) & 3.586 & 13.62 & 23.791 & 33.264  \\
%near user's offloaded time $t_1$ ($\times 10^{-2}$ sec) & 1.82 & 4.73 & 6.32 & 7.01 & 7.46 & 7.8 \\
%far user's offloaded time $t_2$ ($\times 10^{-2}$ sec) &  0.1 & 0.7 & 1.34 & 1.94 & 2.54 & 3.12 \\
Near user's residual energy ($\times 10^{-5}$ Joule)& 0.426 & 3.317 & 6.42 & 9.545  \\
 Far user's residual energy ($\times 10^{-5}$ Joule) & 0 & 0 & 0 & 0 \\
[1ex]
 \hline
\end{tabular}
\end{table*}

Fig.~\ref{fig.perf_vs_d} shows the average energy consumption versus the distance $d_2$ from the AP to the far user, where the computation task sizes for both users are set as $R_1=R_2=20$ kbits. It is observed that as $d_2$ increases, the energy consumption by the six schemes increases significantly, and the proposed joint design achieves the lowest energy consumption among them. The local-computing-only scheme is observed to outperform the full-offloading-only scheme when $d_2<3$ meters, but performs inferior to the full-offloading-only scheme when $d_2>3$ meters. Furthermore, the proposed joint design is observed to achieve a significant performance gain over the separate-design one when $d_2>4$ meters.

To provide more insights, Table~\ref{tab.vs_d} demonstrates the numbers of offloaded bits $\ell_i^{\rm opt}$s at both users and their residual energy (i.e., $E_i-E_{{\rm loc},i}-E_{{\rm offl},i}$) under different values of $d_2$ for the proposed joint design. It is observed that the far user prefers offloading significantly more bits than the near user, especially at a larger $d_2$. As $d_2$ increases, the number of offloaded bits $\ell^{\rm opt}_1$ by the near user decreases significantly, while that by the far user (i.e., $\ell^{\rm opt}_2$) increases. This result is generally consistent with the first property in Remark~\ref{remark1}. Furthermore, it is observed that the residual energy at the near user increases dramatically as $d_2$ increases, while the far user always uses up all its energy when $d_2>d_1$. This shows that as $d_2$ increases, the energy consumption increase at the AP in Fig.~\ref{fig.perf_vs_d} is mainly to satisfy the energy requirement at the far user. In this case, the near user will harvest a lot of energy.

\begin{figure}
  \centering
  \includegraphics[width = 3.5in]{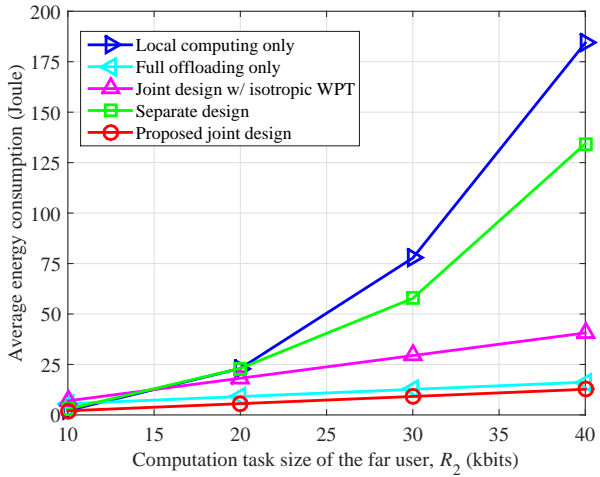}
  \caption{The average energy consumption at the AP versus the computation task size $R_2$ kbits.} \label{fig.perf_vs_size}
\end{figure}

Furthermore, we consider the case when the near and far users have distinct computation task sizes. Fig.~\ref{fig.perf_vs_size} depicts the average energy consumption versus the computation task size $R_2$ at the far user, where $R_1=20$ kbits and $d_2=6$ meters. It is observed that the energy consumption by the six schemes increases as $R_2$ increases, and both the local-computing-only and the separate-design schemes lead to much higher energy consumption than the other four schemes when $R_2>20$ kbits. This is due to the fact that in the local-computing-only and the separate-design schemes, the far user cannot explore the benefit of task offloading for energy saving. Among the five benchmark schemes, the full-offloading-only scheme achieves the best performance close to the optimal proposed one. Table~\ref{tab.vs_size} presents the numbers of offloaded bits at both users and their residual energy for the proposed joint design. It is observed that as $R_2$ increases, the number of offloaded bits $\ell^{\rm opt}_1$ by the near user decreases, while $\ell^{\rm opt}_2$ by the far user increases significantly. Similarly as in Table~\ref{tab.vs_d}, $\ell_2^{\rm opt}$ is observed to be significantly larger than $\ell_1^{\rm opt}$. It is also observed that with $R_2$ increasing, the residual energy at the near user becomes more significant, while that at the far user is zero.

Tables~\ref{tab.vs_d} and \ref{tab.vs_size} show that when the users are heterogeneous in locations and/or task sizes, even the optimal joint design still leads to unbalanced energy demand and supply at these users. In particular, the AP needs to use a large transmit power to satisfy the high energy demand of users that are far apart and/or have heavy computation tasks. At the same time, the nearby users with light computation tasks can accordingly harvest more energy and are likely to have energy surplus. To better balance the energy demand and surplus, it can be viable to enable user cooperation between near and far users, which is an interesting research direction worth pursuing in future work.

\section{Conclusion}
We developed a unified MEC-WPT design framework with joint energy beamforming, offloading, and computing optimization in emerging wireless powered multiuser MEC systems. In particular, we proposed an efficient wireless powered multiuser MEC design by considering the latency-constrained computation, for which the AP minimizes the total energy consumption subject to the users' individual computation latency constrains. Leveraging the Lagrange duality method, we obtained the optimal solution in a semi-closed form. Numerical results demonstrated the merits of the proposed joint design over alternative benchmark schemes. The proposed unified MEC-WPT design can pave the way to facilitate ubiquitous computing for IoT devices in a self-sustainable way.

%\appendices
\section*{Appendices}
\subsection{Proof of Lemma \ref{lem.CPU}}

First, consider the case when there exists some user $i$ with $\ell_i=R_i$, i.e., user $i$ offloads all of its computation task bits to the AP. As user $i$ does not perform local computing in this case, the local CPU frequency of user $i$ is evidently zero.

We next consider the nontrivial case of $0 \leq \ell_i < R_i$, $\forall i\in{\cal K}$. Define $f_i \triangleq \frac{\sum_{n=1}^{C_i(R_i-\ell_i)}f_{i,n}}{C_i(R_i-\ell_i)}$, $\forall i\in{\cal K}$. Since that both ${1}/{x}$ and $x^2$ are convex functions with respect to $x>0$, based on Jensen's inequality\cite{BoydBook}, it follows that
\begin{subequations}\label{eq.C}
\begin{align}
C_i(R_i-\ell_i) /{f_i} &\leq \sum_{n=1}^{C_i(R_i-\ell_i)} {1}/{f_{i,n}} \\
C_i(R_i-\ell_i) \kappa_i f^{2}_i &\leq \sum_{n=1}^{C_i(R_i-\ell_i)}\kappa_i f_{i,n}^2,
\end{align}
\end{subequations}
where both the equalities hold if and only if
\begin{equation}\label{eq.eqn_eqn}
f_{i,1}=\ldots=f_{i,C_i(R_i-\ell_i)},~~\forall i\in{\cal K}.
\end{equation}
As a result, the optimality of problem $({\cal P}1)$ is achieved when (\ref{eq.eqn_eqn}) holds. Therefore, by replacing $f_i\triangleq f_{i,n}$, $\forall n$, problem $({\cal P}1)$ is equivalently expressed as
\begin{subequations}\label{eq.proof_f}
\begin{align}
& \min_{\boldsymbol{Q}\succeq \boldsymbol{0},\boldsymbol{t},\boldsymbol{\ell}, \{f_i\}} ~~ T{\rm tr}(\boldsymbol{Q})+\sum_{i=1}^K \alpha\ell_i \\
  {\rm s.t.} ~~&~~ {C_i(R_i-\ell_i)}/{f_i}  \leq T, ~~\forall i\in{\cal K} \\
 &~~ \kappa_iC_i(R_i-\ell_i)f_i^2 +\frac{t_i}{\tilde{g}_i}\beta\left(\frac{\ell_i}{t_i}\right) +p_{c,i} t_i \notag \\
 &~~~~~~~~~~~~~~~~~~~~ - T\zeta{\rm tr}(\boldsymbol{Q}\boldsymbol{H}_i) \leq 0, ~~\forall i \in{\cal K}\\
 & ~~\sum_{i=1}^K t_i \leq T,~~t_i\geq 0,~~0 \leq \ell_i \leq R_i,~~\forall i \in{\cal K}.
\end{align}
\end{subequations}
For a given $(\boldsymbol{t},\boldsymbol{\ell})$, it is evident that the optimal $f_i$'s for (\ref{eq.proof_f}) (equivalent $({\cal P}1)$) should be as small as possible by (\ref{eq.proof_f}c). Since $f_i$ is bounded below by ${C_i(R_i-\ell_i)}/{T}$ in (\ref{eq.proof_f}b), it follows that the optimal $f_i$'s are
\begin{equation}
f_i = {C_i(R_i-\ell_i)}/{T},~~\forall i \in{\cal K}.
\end{equation}
It then readily follows that, at optimum of $({\cal P}1)$, $
f_{i,1} =...=f_{i,C_i(R_i-\ell_i)} = {C_i(R_i-\ell_i)}/{T}$, $\forall i \in{\cal K}
$.

\subsection{Proof of $\boldsymbol{F}(\boldsymbol{\lambda})\succeq \boldsymbol{0}$}

$\boldsymbol{F}(\boldsymbol{\lambda})\succeq \boldsymbol{0}$ can be verified by contradiction. Assume that $\boldsymbol{F}(\boldsymbol{\lambda})$ is not positive semidefinite. Denote by $\boldsymbol{\xi}$ one eigenvector corresponding to the negative eigenvalue of $\boldsymbol{F}(\boldsymbol{\lambda})$. By setting $\boldsymbol{Q}=\tau \boldsymbol{\xi}\boldsymbol{\xi}^H\succeq \boldsymbol{0}$ with $\tau$ going to infinity (which is feasible for (\ref{eq.dual_obj1})), it follows that
\begin{equation}
\lim_{\tau \rightarrow +\infty}~{\rm tr}(\boldsymbol{Q}\boldsymbol{F}(\boldsymbol{\lambda})) = \lim_{\tau \rightarrow +\infty} ~\tau\boldsymbol{\xi}^H \boldsymbol{F}(\boldsymbol{\lambda})\boldsymbol{\xi} = -\infty,
\end{equation}
which in turn implies that the objective value in (\ref{eq.dual_obj1}) is unbounded below over $\boldsymbol{Q}\succeq \boldsymbol{0}$. Therefore, in order for the dual function value $\Phi(\boldsymbol{\lambda},\mu)$ to be bounded below, we need $\boldsymbol{F}(\boldsymbol{\lambda})\succeq \boldsymbol{0}$.

\subsection{Proof of Lemma \ref{lem.t_l_sol1}}
Given $(\boldsymbol{\lambda},\mu)\in {\cal S}$, we solve problem (\ref{eq.La1_dist}) for each user $i\in{\cal K}$. When $\lambda_i=0$, the objective function in (\ref{eq.La1_dist}) becomes $\alpha \ell_i + \mu t_i$. It is evident that $t^*_i=0$ and $\ell^*_i=0$ are optimal for (\ref{eq.La1_dist}).

For $\lambda_i>0$, the Lagrangian of (\ref{eq.La1_dist}) is given by
\begin{align}
{\cal L}_i  =~& \alpha \ell_i+\frac{\lambda_i \kappa_i C^3_i(R_i-\ell_i)^3}{T^2}+\frac{\lambda_it_i}{\tilde{g}_i}\beta\left(\frac{\ell_i}{t_i}\right)+\lambda_i p_{c,i}t_i \notag \\
&+ \mu t_i + \gamma_i(\ell_i-R_i)-\nu_i \ell_i -\eta_i t_i,
\end{align}
where $\gamma_i$, $\nu_i$, and $\eta_i$ are the non-negative Lagrangian multipliers associated with $\ell_i\leq R_i$, $\ell_i\geq 0$, and $t_i\geq 0$, respectively. Based on the KKT conditions\cite{BoydBook}, the necessary and sufficient conditions for the optimal primal-dual point $(t^*_i, \ell^*_i , \gamma^*_i, \nu^*_i, \eta^*_i)$ are
\begin{subequations}\label{eq.kkt}
\begin{align}
& t^*_i\geq 0,~~0\leq \ell^*_i \leq R_i\\
& \gamma^*_i\geq 0,~~\nu^*_i \geq 0,~~\eta^*_i \geq 0\\
&  \gamma^*_i(\ell_i^*-R_i) = 0,~~\nu^*_i \ell^*_i = 0, ~~\eta^*_i t^*_i=0\\
&\frac{\lambda_i}{\tilde{g}_i} \left( \beta\left(\frac{\ell^*_i}{t^*_i}\right)-\frac{\ell^*_i}{t^*_i}\beta'\left(\frac{\ell^*_i}{t^*_i}\right)\right)+\lambda_ip_{c,i}+\mu-\eta^*_i=0\\
&\alpha - \frac{3\lambda_i\kappa_iC^3_i(R_i-\ell^*_i)^2}{T^2}+\frac{\lambda_i}{\tilde{g}_i}\beta'\left(\frac{\ell^*_i}{t^*_i}\right)+\gamma^*_i-\nu^*_i = 0,
\end{align}
\end{subequations}
where $\beta'(x)\triangleq \frac{\sigma^2\ln 2}{B}2^{\frac{x}{B}}$ is the first-order derivative of $\beta(x)$ with respect to $x$. Note that (\ref{eq.kkt}c) denotes the complementary slackness condition, while the left-hand-side terms of (\ref{eq.kkt}d) and (\ref{eq.kkt}e) are the first-order derivatives of ${\cal L}_i$ with respect to $t^*_i$ and $\ell^*_i$, respectively.
For the function $y=\beta(x)-x\beta'(x)$ of $x>0$, its inverse function can be shown to be \cite{LambertW}
\begin{equation}\label{eq.sol_W0}
x = \frac{B}{\ln 2}\left( W_0\left(-\frac{y}{\sigma^2 e}-\frac{1}{e}\right) +1\right).
\end{equation}
Let $r_i^*\triangleq {\ell_i^*}/{t_i^*}$. From (\ref{eq.kkt}b) and (\ref{eq.kkt}d), we have
$\beta(r_i^*)-r_i^*\beta'(r_i^*) = -\tilde{g}_i \left(\frac{\mu}{\lambda_i}+p_{c,i}\right)$. Based on (\ref{eq.sol_W0}), it follows that
\begin{equation}\label{eq.r_sol}
r^*_i = \frac{B}{\ln 2}\left(W_0  \left(\frac{\tilde{g}_i}{\sigma^2 e}\left( \frac{\mu}{\lambda_i}+p_{c,i} \right)-\frac{1}{e}\right)+ 1\right).
\end{equation}
Since $W_0(x)$ is a monotonically increasing function of $x\geq -\frac{1}{e}$ and $W_0(-\frac{1}{e})=-1$\cite{LambertW}, it follows that $r_i^*>0$ with non-zero $p_{{\rm c},i}$. From (\ref{eq.kkt}c) and (\ref{eq.kkt}e), it is immediate that
\begin{equation} \label{eq.l_sol}
\ell^*_i = \left[ R_i- \sqrt{\frac{1}{3\kappa_i{C}^3_i} \left( \frac{\alpha}{\lambda_i} +\frac{\sigma^2 \ln 2}{B\tilde{g}_i}2^{\frac{r^*_i}{B}} \right) } \right]^+.
\end{equation}
With (\ref{eq.r_sol}) and (\ref{eq.l_sol}), the optimal $t_i^*$ is then obtained as $t^*_i =\ell^*_i/r^*_i$.

\subsection{Proof of Lemma \ref{lem.F_sub}}
The positive semidefinite constraint $\boldsymbol{F}(\boldsymbol{\lambda})\succeq \boldsymbol{0}$ can be equivalently expressed as a scalar inequality constraint as\cite{BoydBook}
\begin{equation}\label{eq.pi}
\pi(\boldsymbol{\lambda})\triangleq \min_{\|\boldsymbol{\xi}\|=1}\boldsymbol{\xi}^H\boldsymbol{F}(\boldsymbol{\lambda})\boldsymbol{\xi} \geq 0.
\end{equation}
Given a query point $\boldsymbol{\lambda}_1\triangleq [\lambda_{1,1},\ldots,\lambda_{1,K}]^\dagger$, one can find the normalized eigenvector $\boldsymbol{v}_1$ of $\boldsymbol{F}(\boldsymbol{\lambda}_1)$ corresponding to the smallest eigenvalue of ${\bm F}({\bm \lambda}_1)$ (i.e., $\pi(\boldsymbol{\lambda}_1)$). Consequently, we can determine the value of the scalar constraint at a query point as $\pi(\boldsymbol{\lambda}_1)=\boldsymbol{v}_1^H\boldsymbol{F}(\boldsymbol{\lambda}_1)\boldsymbol{v}_1$. To obtain a subgradient, we have
\begin{subequations}
\begin{align}
\pi(\boldsymbol{\lambda}) -\pi(\boldsymbol{\lambda}_1)
&= \min_{\|\boldsymbol{\xi}\|=1} \boldsymbol{\xi}^H\boldsymbol{F}(\boldsymbol{\lambda}) \boldsymbol{\xi} - \boldsymbol{v}^H_1\boldsymbol{F}(\boldsymbol{\lambda}_1)\boldsymbol{v}_1 \notag \\
& \leq \boldsymbol{v}^H_1\left( \boldsymbol{F}(\boldsymbol{\lambda}) - \boldsymbol{F}(\boldsymbol{\lambda}_1)\right)\boldsymbol{v}_1 \\
& = \sum_{i=1}^K \left(  \lambda_{1,i}  - \lambda_{i}\right) \zeta\boldsymbol{v}^H_1\boldsymbol{H}_i \boldsymbol{v}_1,
\end{align}
\end{subequations}
where the last equality follows from the affine structure of $\boldsymbol{F}(\cdot)$ in (\ref{eq.dual_prob1}b). By the weak subgradient calculus\cite{BoydBook}, the subgradient of $\boldsymbol{F}(\boldsymbol{\lambda})$ at the given $\boldsymbol{\lambda}$ and $\mu$ is then
\begin{equation}
[\zeta\boldsymbol{v}^H\boldsymbol{H}_1\boldsymbol{v},\ldots,\zeta\boldsymbol{v}^H\boldsymbol{H}_K\boldsymbol{v}, 0]^\dagger,
\end{equation}
where $\boldsymbol{v}$ is the eigenvector corresponding to the smallest eigenvalue of $\boldsymbol{F}(\boldsymbol{\lambda})$, and the last zero entry follows from the fact that $\pi(\boldsymbol{\lambda})$ is independent of $\mu$.

\ifCLASSOPTIONcaptionsoff
  \newpage
\fi

\begin{IEEEbiography}[{\includegraphics[width=1in,height=1.25in,clip,keepaspectratio]{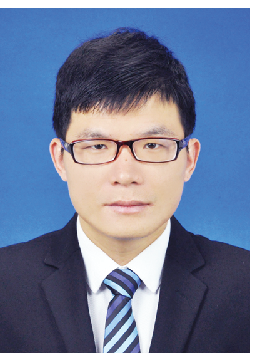}}]
{Feng Wang}(M'16) received the B.Eng. degree from Nanjing University of Posts and Telecommunications, China, in 2009, and the M.Sc. and Ph.D. degrees, both from Fudan University, China, in 2012 and 2016, respectively. He is currently an Assistant Professor with the School of Information Engineering, Guangdong University of Technology, China. From 2012 to 2013, he was a Research Fellow with the Department of Communication Technology, Sharp Laboratories of China. From January 2017 to September 2017, he was a Postdoctoral Research Fellow with the Engineering Systems and Design Pillar, Singapore University of Technology and Design. His research interests include signal processing for communications, wireless power transfer, and edge computing.
\end{IEEEbiography}

\begin{IEEEbiography}[{\includegraphics[width=1in,height=1.25in,clip,keepaspectratio]{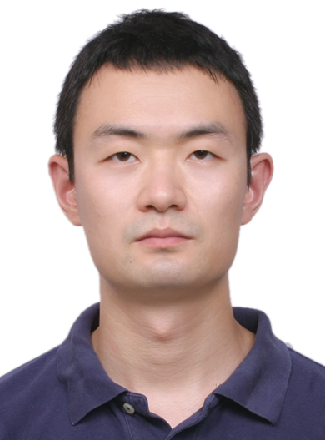}}]{Jie Xu}(S'12-M'13) received his B.E. and Ph.D. degrees from University of Science and Technology of China in 2007 and 2012, respectively. He is now with the School of Information Engineering, Guangdong University of Technology, China. From 2012 to 2014, he was a Research Fellow with the Department of Electrical and Computer Engineering, National University of Singapore. From 2015 to 2016, he was a Postdoctoral Research Fellow with the Engineering Systems and Design Pillar, Singapore University of Technology and Design. His research interests include energy efficiency and energy harvesting in wireless communications, wireless information and power transfer, wireless security, UAV communications, and mobile edge computing. He is now an Editor for the IEEE WIRELESS COMMUNICATIONS LETTERS, an Associate Editor for the IEEE ACCESS, and a Guest Editor for the IEEE WIRELESS COMMUNICATIONS. He received the IEEE Signal Processing Society Young Author Best Paper Award in 2017.
\end{IEEEbiography}

\begin{IEEEbiography}[{\includegraphics[width=1in,height=1.25in,clip,keepaspectratio]{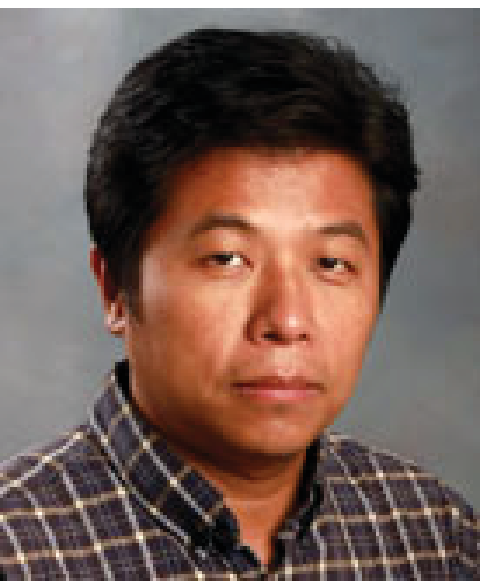}}]
{Xin Wang} (SM'09) received the B.Sc. and M.Sc. degrees from Fudan University, Shanghai, China, in 1997 and 2000, respectively, and the Ph.D. degree from Auburn University, Auburn, AL, USA, in 2004, all in electrical engineering.

From September 2004 to August 2006, he was a Postdoctoral Research Associate with the Department of Electrical and Computer Engineering, University of Minnesota, Minneapolis. In August 2006, he joined the Department of Computer and Electrical Engineering and Computer Science, Florida Atlantic University, Boca Raton, FL, USA, as an Assistant Professor, and then an Associate Professor from August 2010. He is currently a Distinguished Professor with the Department of Communication Science and Engineering, Fudan University. His research interests include stochastic network optimization, energy-efficient communications, cross-layer design, and signal processing for communications. He served as an Associate Editor for the IEEE Signal Processing Letters. He currently serves as an Associate Editor for the IEEE Transactions on Signal Processing and as an Editor for the IEEE Transactions on Vehicular Technology.
\end{IEEEbiography}

\begin{IEEEbiography}[{\includegraphics[width=1in,height=1.25in,clip,keepaspectratio]{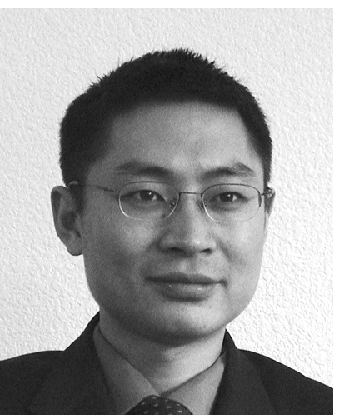}}]
{Shuguang Cui} (S'99-M'05-SM'12-F'14) received his Ph.D in Electrical Engineering from Stanford University, California, USA, in 2005. Afterwards, he has been working as assistant, associate, and full professor in Electrical and Computer Engineering at the Univ. of Arizona and Texas A\&M University. He is currently a Child Family Endowed Chair Professor in Electrical and Computer Engineering at the Univ. of California-Davis. His current research interests focus on data driven large-scale system control and resource management, large data set analysis, IoT system design, energy harvesting based communication system design, and cognitive network optimization. He was selected as the Thomson Reuters Highly Cited Researcher and listed in the Worlds' Most Influential Scientific Minds by ScienceWatch in 2014. He was the recipient of the IEEE Signal Processing Society 2012 Best Paper Award. He has served as the general co-chair and TPC co-chairs for many IEEE conferences. He has also been serving as the area editor for IEEE Signal Processing Magazine, and associate editors for IEEE Transactions on Big Data, IEEE Transactions on Signal Processing, IEEE JSAC Series on Green Communications and Networking, and IEEE Transactions on Wireless Communications. He has been the elected member for IEEE Signal Processing Society SPCOM Technical Committee (2009$\sim$2014) and the elected Chair for IEEE ComSoc Wireless Technical Committee (2017$\sim$2018). He is a member of the Steering Committee for both IEEE Transactions on Big Data and IEEE Transactions on Cognitive Communications and Networking. He is also a member of the IEEE ComSoc Emerging Technology Committee. He was elected as an IEEE Fellow in 2013 and an IEEE ComSoc Distinguished Lecturer in 2014.
\end{IEEEbiography}

\end{document}